\documentclass[letterpaper]{article} 
\usepackage[]{aaai25}  
\usepackage{times}  
\usepackage{helvet}  
\usepackage{courier}  
\usepackage[hyphens]{url}  
\usepackage{graphicx} 
\usepackage{natbib}  
\usepackage{caption} 
\usepackage{algorithm}
\usepackage{algorithmic}

\usepackage{newfloat}
\usepackage{listings}
\DeclareCaptionStyle{ruled}{labelfont=normalfont,labelsep=colon,strut=off} 
\floatstyle{ruled}
\newfloat{listing}{tb}{lst}{}
\floatname{listing}{Listing}
\nocopyright

\usepackage{booktabs}
\usepackage{comment}
\usepackage{amsmath}

\usepackage{xcolor}

\title{Converging Naming Styles, Persistent Network Locality: GitHub in the LLM Era}
\author {
    Yuto Tamura\textsuperscript{\rm 1},
    Sho Tsugawa\textsuperscript{\rm 2},
}
\affiliations {
    \textsuperscript{\rm 1} Doctoral Program in Computer Science, Graduate School of Science and Technology, Univerisity of Tsukuba\\
    \textsuperscript{\rm 2}Institute of Systems and Information Engineering, University of Tsukuba\\
    yuto.tamura@snlab.cs.tsukuba.ac.jp, s-tugawa@cs.tsukuba.ac.jp
}

\begin{document}

\maketitle

\begin{abstract}
Social conventions often emerge through repeated interactions within social networks, allowing shared practices to coexist with variation across groups. Large language models (LLMs) introduce a potentially different coordination structure: a small number of widely used models can expose socially distant users to similar patterns and suggestions. Whether broad convergence under such shared technological influences eliminates network-local variation remains unclear.
We examine this question in software development, where identifier naming styles provide observable conventions and LLM-based tools have rapidly diffused. Using public GitHub repositories created between 2015 and September 2025 across six programming languages, we characterize naming styles with 27 features and examine their association with detected LLM-related commits, their diversity across repository creation cohorts, and their relationship to owner proximity in a large-scale collaboration network.
Repositories with detected LLM-related commits tend to use longer identifiers and, in several languages, make greater use of naming patterns already prevalent within the language. We also observe lower naming-style diversity in recent creation cohorts, with marked declines appearing around 2023--2024 in several languages, although their timing and trajectories differ. At the same time, network locality persists: in five of the six languages, repositories whose owners are closer in the collaboration network remain more similar in naming style even among recent, more homogeneous cohorts.
These findings show that aggregate convergence and network-local variation can coexist, highlighting the need to examine not only how much cultural variation remains, but also how that variation continues to be structured by human social relationships in the era of widely shared AI systems.
\end{abstract}

\section{Introduction}
Social conventions are shaped through repeated interactions among people engaged in shared activities~\cite{Young_evoluoin_of_convention,Centola_Baronchelli_cordination_structure}. By aligning their practices, people develop common ways of doing things that facilitate coordination. Because these interactions are structured by social networks, such alignment is often local: socially proximate individuals tend to share conventions, while different groups may sustain distinct practices~\cite{Axelrod_local_convention_global_pol,Centola_long_tie,community_toxic_norm,Khalid_2020_style_community}. At the same time, conventions can spread beyond particular groups and become widely shared~\cite{Centola_Baronchelli_cordination_structure}. The distribution of social conventions thus encompasses both broad commonalities and local variation, reflecting the interplay between the wider diffusion of practices and coordination within social networks.

Large language models (LLMs) introduce a potentially population-scale channel through which conventions may become shared~\cite{brinkmann2023machine_culture,wikipedia_norms}. A relatively small number of widely used models interact with many users, potentially exposing people who are distant in social networks to similar patterns and practices. This shared source of suggestions could promote convergence without requiring conventions to spread through interpersonal ties. Yet greater overall similarity need not eliminate differences associated with human relationships: local coordination may continue alongside influences from common models. The emergence and diffusion of LLMs therefore raise an important question about the distribution of social conventions: does broader convergence diminish network-local variation, or can the two coexist?

Software development on GitHub provides a concrete setting for examining this question. LLM-based tools have become increasingly integrated into programming practices~\cite{vibecoding_survey}, while software development also involves coordination among human collaborators~\cite{Lima_coding_togther_at_scale}. We focus on identifier naming as an observable aspect of coding conventions, including the length~\cite{Lawrie_naming_style_length,Binkley_identifier_name} and composition of names and the use of naming patterns such as \texttt{snake\_case} and \texttt{camelCase}~\cite{Binkley_camel_underscore}. These features provide measurable indicators of practices that can vary across projects within the same programming language. Public source code makes this variation observable, while collaborative activity records allow naming-style similarity to be examined in relation to the network proximity of repository owners. This setting therefore enables us to study overall diversity and network locality within the same domain of conventions.

In this study, we investigate the relationship between broad convergence and network locality in identifier naming styles in the context of LLM diffusion. Our central interest is whether greater overall similarity coexists with persistent variation associated with collaboration-network proximity. We examine associations with detected LLM-related commits and compare repositories grouped by their creation periods, which we refer to as creation cohorts. Specifically, we address the following research questions:

\begin{itemize}
    \item \textbf{RQ1:} How do identifier naming styles differ between repositories with and without detected LLM-related commits?
    \item \textbf{RQ2:} How does naming-style diversity vary across repository creation cohorts before and during the diffusion of LLM-based development tools?
    \item \textbf{RQ3:} How does collaboration-network proximity relate to naming-style similarity, and how does this relationship vary across repository creation cohorts?
\end{itemize}

To address these questions, we analyze public GitHub repositories created between January 2015 and September 2025 across six programming languages. We represent each repository's identifier naming style using 27 features extracted from the default-branch code available at clone time. Our temporal comparisons are therefore organized by repository creation cohort rather than based on reconstructed historical code snapshots. We compare naming features between repositories with and without detected LLM-related commits and quantify within-cohort diversity using pairwise distances between naming-style feature vectors. We further examine how these distances relate to repository owners' proximity in a fixed collaboration network linking users who pushed to the same repository. These observational comparisons characterize associations with LLM-related development activity and repository creation periods, rather than identify causal effects of LLM use.

Our contributions are threefold:
\begin{itemize}
    \item We characterize naming-style differences associated with detected LLM-related commits. Repositories with such commits tend to use longer identifiers and, in several languages, make greater use of naming patterns already prevalent within the language.

    \item We document a broad decline in naming-style diversity across recent repository creation cohorts. Declines become apparent around 2023--2024 in several languages, although their timing and magnitude vary across languages.

    \item We show that overall homogenization coexists with network-local variation. In most of the studied languages, repositories whose owners are closer in the collaboration network exhibit more similar naming styles, and this association persists in recent creation cohorts despite their lower overall diversity.
\end{itemize}

\section{Related Work}

A substantial body of research has examined how social conventions emerge, spread, and remain differentiated across social networks~\cite{Centola_Baronchelli_cordination_structure,Rotabi_2017,Kooti_Emergence_of_osn2012}.
\citet{Centola_Baronchelli_cordination_structure} experimentally investigated whether local coordination among network neighbors can give rise to population-wide conventions, showing that network structure can lead either to global convergence or to the persistence of multiple local conventions.
\citet{Rotabi_2017} examined how practices are transmitted through collaboration networks, using scientific writing as an empirical setting.
\citet{Kooti_Emergence_of_osn2012} studied competing conventions for representing reposts in an online social network and characterized their adoption dynamics as diffusion over the follower network.
Together, these studies show that the distribution of conventions depends not only on their intrinsic properties, but also on the network structure through which people coordinate and observe one another.

In software engineering, coding styles and identifier naming practices have similarly been studied as conventions shared within projects and development communities.
\citet{Allamanis_2014} characterized project-specific coding conventions as an emergent consensus arising from developers' local stylistic decisions.
Empirical studies have identified systematic and project-specific patterns in identifier naming~\cite{Butler_2011}, while substantial variation has also been observed in the names chosen by individual developers~\cite{Feitelson_2022}.
Moreover, identifier styles evolve over the lifetime of software projects~\cite{Zhang_2021}.
These findings suggest that identifier naming provides a useful observable form of convention: naming choices vary across developers and projects, yet become structured and shared within development communities.

Recent work has begun to examine how large language models may reshape the distribution of linguistic and cultural outputs.
From a broader perspective, the concept of machine culture highlights the possibility that intelligent systems participate in cultural variation, transmission, and selection~\cite{brinkmann2023machine_culture}.
Empirically, \citet{Doshi_Hauser_diverisity} found that access to LLM-generated suggestions improved evaluations of individual written outputs while making the resulting works more similar to one another, reducing collective diversity.
\citet{Acerbi_Stubbersfield} showed that large language models exhibit content biases similar to those observed in humans when mediating information transmission.
At the population level, \citet{Kodak_delving} documented sharp increases in the frequency of particular words, such as \textit{delve}, in biomedical abstracts during the period of widespread LLM-assisted writing.
These studies suggest that widely shared generative systems may affect not only individual outputs but also the distribution of conventions and stylistic choices across a population.

A related line of work has examined the stylistic and quality characteristics of code generated by large language models.
\citet{AlMadi2022Copilot} compared GitHub Copilot-generated code with human-written code and found broadly comparable readability and complexity.
\citet{Liu2024ChatGPTCodeQuality} analyzed 4,066 programs generated by ChatGPT and identified a range of quality issues, including problems related to coding style and maintainability.
More directly focusing on stylistic differences, \citet{Wang2025CodingStyle} compared code generated by five LLMs with human-written code and identified 24 types of coding-style inconsistencies spanning formatting, naming, expressions and statements, control flow, and fault tolerance.
They further found that coding styles were relatively similar across different LLMs while systematically differing from human-written code.
This literature establishes that LLM-generated code can exhibit characteristic stylistic patterns, but it primarily compares generated and human-written code rather than examining how conventions are distributed across large populations of real-world projects and collaboration networks.

Closest to our work, \citet{xu-etal-2026-code} examined both LLM-based code transformation and temporal variation in real-world source code.
In controlled experiments, they showed that code rewritten by LLMs using human-written code as a reference remained substantially more similar to the original code than code generated from scratch.
Separately, they analyzed more than 20,000 GitHub repositories linked to arXiv papers, focusing on Python and C/C++, and documented temporal shifts in coding styles, including increased use of \texttt{snake\_case} function names in Python.
Their study provides important evidence that coding-style distributions have changed during the period in which LLM-based development tools became widespread.

Our study complements and extends this literature by focusing on both the overall distribution of identifier naming styles and its organization across a collaboration network.
First, we examine repositories across six programming languages and are not restricted to projects associated with a particular research domain or application area.
Second, through retrospective comparisons across repository creation cohorts, we quantify how within-language diversity in identifier naming styles differs across periods before and during the diffusion of LLM-based development tools.
Third, and most importantly, we relate naming-style similarity to distance in a large-scale developer collaboration network and examine whether network-local variation persists as overall naming-style diversity declines.
This joint analysis of aggregate convergence and network locality allows us to ask whether increasingly shared conventions necessarily erase differences structured by human collaboration networks.



\section{Data and Methods}
\label{sec:data_methods}

\subsection{Data Collection and Sampling}
\label{sec:data_collection}

We analyzed source code from public GitHub repositories created between January 2015 and September 2025 across six programming languages: Python, JavaScript, TypeScript, Java, Go, and Rust. We retrieved repository metadata through the GitHub API using the \texttt{language:} and \texttt{created:} search qualifiers to specify programming language and repository creation date, respectively. The corresponding metadata fields were \texttt{primaryLanguage.name} and \texttt{createdAt}. We then filtered the returned metadata using \texttt{diskUsage}, \texttt{isFork}, and \texttt{createdAt}, retaining non-fork repositories created within the specified period whose metadata-reported sizes were strictly greater than 10~KB and strictly less than 5{,}000~KB. Repositories were subsequently sampled from the filtered candidates according to the procedures described below and cloned for source-code analysis.

We used three analysis-specific sampling schemes. For \textbf{RQ1}, we selected repositories created in September 2025 and divided them into two groups according to whether our search detected at least one LLM-related commit. For this comparison, the detection window covered September--October 2025; the search and classification procedures are described in Section~\ref{sec:llm_detection}. We randomly sampled up to 2{,}000 repositories per language from each group. The comparison therefore distinguishes repositories with and without detected LLM-related commits, rather than verified users and non-users of LLM-based tools.

For \textbf{RQ2}, we randomly sampled up to 2{,}000 repositories per language and creation month over the period from January 2015 to September 2025. We applied the same metadata-based filtering criteria as for RQ1 and sampled without replacement using a fixed random seed. Within this overall period, sampling began in January 2017 for TypeScript and January 2020 for Rust because insufficient repositories were available for sampling in earlier periods; sampling for the other four languages began in January 2015. Only successfully cloned repositories were retained for subsequent analysis, so the number retained for a given language and month could be fewer than 2{,}000. We subsequently grouped these repositories by creation quarter to compare naming-style diversity across creation cohorts. Each cohort thus contains repositories created within the same calendar quarter, rather than repeated observations of the same repositories.

For \textbf{RQ3}, we sampled repositories whose owners belonged to the largest connected component (LCC) of the collaboration network described in Section~\ref{sec:network_locality}. Sampling was performed by language and repository creation month. We sampled up to 5{,}000 repositories per month for Python, Java, and JavaScript from January 2015 to September 2025; up to 5{,}000 per month for TypeScript and 2{,}000 per month for Go from January 2018 to September 2025; and up to 1{,}000 per month for Rust from January 2020 to September 2025. As in RQ2, the sampled repositories were grouped by creation quarter.

Across all three samples, we analyzed the default-branch HEAD available when each repository was cloned. Source-code collection took place on September 14, 2026, for RQ1; from July 29, 2025, to January 25, 2026, for RQ2; and from April 22 to April 29, 2026, for RQ3. Each repository was cloned at least one month after its creation. We did not use \texttt{git checkout} to retrieve historical snapshots corresponding to repository creation dates. Accordingly, repository creation time defines cohort membership, whereas the analyzed source code reflects the snapshot obtained at collection.

To assess the correspondence between repository creation time and the timing of the observed code, we compared repository creation timestamps with the author timestamps of the HEAD commits recorded in the saved clones. Among repositories created before January 2025, the cumulative proportion at day 30 was 69.4\% for the RQ2 clone sample and 70.6\% for the RQ3 clone sample. These cumulative proportions include negative intervals. Appendix~\ref{app:repository_creation_and_observed_code} presents the distributions of elapsed time between these timestamps. Because Git author timestamps can be set arbitrarily when commits are created or history is rewritten, they may not reflect the actual last-modification times of the analyzed source files. These summaries therefore provide only an approximate assessment of temporal alignment and do not replace historical snapshots.

Table~\ref{tab:sample_sizes} summarizes the final numbers of repositories included in each analysis. Repositories were excluded if cloning failed, naming features were unavailable, or no identifiers were extracted in at least one of the three categories: functions, parameters, and variables. Collection and processing coverage, including the numbers retained by language and creation quarter, is reported in Appendix~\ref{app:sample_coverage}. The final RQ3 sample contains 1{,}237{,}319 distinct repository owners across all languages and creation cohorts. Owner counts by language and creation quarter are also reported in that appendix.

\begin{table}[t]
    \centering
    \small
    \setlength{\tabcolsep}{4pt}
    \begin{tabular}{lrrrr}
    \hline
    & \multicolumn{2}{c}{RQ1} & RQ2 & RQ3 \\
    Language & Detected & Not detected & Cohort & Network \\
    \hline
    Python & 1,644 & 1,572 & 225,276 & 574,575 \\
    JavaScript & 1,342 & 828 & 171,067 & 384,484 \\
    TypeScript & 1,206 & 957 & 76,688 & 179,876 \\
    Java & 1,032 & 1,773 & 247,191 & 614,245 \\
    Go & 734 & 1,700 & 244,642 & 171,738 \\
    Rust & 745 & 1,713 & 128,832 & 62,226 \\
    \hline
\end{tabular}
    \caption{
        Final numbers of repositories included in each analysis.
        For RQ1, ``Detected'' and ``Not detected'' indicate whether
        our search identified LLM-related commits.
        ``Cohort'' and ``Network'' refer to the RQ2 creation-cohort
        sample and the RQ3 network-restricted sample, respectively.
    }
    \label{tab:sample_sizes}
\end{table}

\subsection{Identifying LLM-Related Commits}
\label{sec:llm_detection}

 We identified LLM-related commits through the GitHub commit search API. We issued separate, case-insensitive keyword searches for \texttt{ChatGPT}, \texttt{Claude}, \texttt{Codex}, \texttt{Copilot}, and \texttt{DeepSeek} in commit messages on public repositories' default branches. A match to any of these keywords was sufficient for detection. The search window extended from January 1, 2015, through November 1, 2025, using the \texttt{author-date} qualifier to filter commits. Returned records were linked to repository metadata by matching \texttt{repository\_node\_id} in the collected commit records to \texttt{node\_id} in the repository metadata.

For the RQ1 comparison, repositories created in September 2025 were classified using all commit records in the five collected keyword-specific datasets. A repository was assigned to the detected group when at least one retrieved commit matched a keyword; otherwise, it was assigned to the not-detected group. We did not apply additional exclusions for bot-authored commits, merge commits, or reverts. Commit records were not deduplicated across repositories. Repeated matches within a repository did not change its binary label, whereas matches associated with different repositories were retained for their respective classifications.

This classification identifies observed references to LLM-related tools, not verified use of those tools to generate the analyzed code. A matching message may refer to LLM-related functionality rather than coding assistance, and the absence of a detected message does not establish non-use. We therefore use the terms ``detected'' and ``not detected'' throughout the analysis.

Among the collected candidate repositories satisfying the language, non-fork, and size criteria and created in September 2025, the proportions of repositories with detected LLM-related commits were 4.0\% for Python, 1.7\% for JavaScript, 3.9\% for TypeScript, 0.80\% for Java, 4.6\% for Go, and 6.1\% for Rust.

A descriptive overview of the retrieved LLM-related commit records, including monthly counts by keyword, is provided in Appendix~\ref{app:llm_commit_records}.

\subsection{Measuring Identifier Naming Style}
\label{sec:naming_features}

We characterized naming style using identifiers extracted from source code. For each repository, we selected files with the extension corresponding to its sampled primary language: \texttt{.py}, \texttt{.js}, \texttt{.ts}, \texttt{.java}, \texttt{.go}, or \texttt{.rs}. Files in other languages were not included. We excluded version-control directories and designated dependency, build, cache, and generated-code directories using language-specific directory-name filters. We parsed the selected files using Tree-sitter 0.25.2 with tree-sitter-language-pack 0.11.0 and extracted function names, parameter names, and variable names.

Identifiers were counted once per name per file within each category, even if the same name was defined multiple times in that file; the same name appearing in different files was counted separately. For each of the three identifier categories, we pooled the names across files within each repository and calculated nine features from the pooled names: the mean number of characters~\cite{Lawrie_naming_style_length,Hofmeister_2018}; the mean number of constituent words; and the proportions of identifiers consisting of a single word, consisting of one character, consisting of two characters~\cite{Lawrie_naming_style_length}, using \texttt{snake\_case}, using \texttt{camelCase}~\cite{Binkley_camel_underscore}, consisting entirely of uppercase letters, and containing digits. Together, these features yielded a 27-dimensional vector
$\mathbf{x}_r=(x_{r1},\ldots,x_{r27})$ for repository $r$.

For word counting, identifiers were split at non-alphanumeric separators, lowercase-to-uppercase and digit-to-uppercase transitions, and acronym-to-word boundaries. Only segments containing alphabetic characters counted as words; leading and trailing separators were ignored, and digits attached to words added no words. Naming-pattern indicators could overlap: snake case required at least three characters and an underscore remaining after trimming leading and trailing underscores; camel case required both lowercase and uppercase characters; and uppercase required at least one alphabetic character, with all alphabetic characters uppercase regardless of digits or symbols. Repositories with no extracted function, parameter, or variable names in any one category were excluded from the analyses.

\subsection{Comparing Naming Styles by LLM-Related Commit Detection}
\label{sec:rq1_comparison}

To address RQ1, we compared each naming feature between the detected and not-detected repository groups, separately for each programming language. The unit of comparison was the repository. Comparisons used untransformed feature values, without clipping or standardization. Descriptive group means were calculated on the original feature scales, giving equal weight to each repository.

We summarized the direction and magnitude of pairwise ordering using the Mann--Whitney AUC. For a given language and feature $j$, let $\mathcal{R}_{D}$ and $\mathcal{R}_{N}$ denote the repositories included in the detected and not-detected comparisons, with sizes $n_D$ and $n_N$. The AUC is
\begin{equation}
    A_j =
    \frac{1}{n_D n_N}
    \sum_{r\in\mathcal{R}_{D}}
    \sum_{s\in\mathcal{R}_{N}}
    H(x_{rj},x_{sj}),
    \label{eq:rq1_auc}
\end{equation}
where
\begin{equation}
    H(a,b) =
    \mathbf{1}\{a>b\}
    + \frac{1}{2}\mathbf{1}\{a=b\},
    \label{eq:auc_comparison}
\end{equation}
and $\mathbf{1}\{\cdot\}$ is the indicator function. Values above 0.5 indicate a tendency toward higher feature values in the detected group, whereas values below 0.5 indicate the opposite tendency. A value of 0.5 indicates no directional ordering under this comparison, rather than necessarily identical feature distributions.

\subsection{Measuring Naming-Style Diversity across Creation Cohorts}
\label{sec:cohort_diversity}

To address RQ2, we measured naming-style diversity as the mean pairwise distance between repositories in the same language and creation quarter. Before computing distances, we normalized the 27 features separately for each language.   Transformation parameters were fitted using all repositories retained for the RQ2 analysis in each language, pooled across all included creation cohorts, and were held fixed across quarters.s

For feature $j$ in language $\ell$, let $a_{\ell j}$ and $b_{\ell j}$ denote its 1st and 99th percentiles in the reference sample. We clipped feature values at these thresholds and applied min--max scaling. For $b_{\ell j}>a_{\ell j}$, the normalized value was
\begin{equation}
    z_{rj}^{(2)}
    =
    \frac{
        \min\{\max\{x_{rj},a_{\ell j}\},b_{\ell j}\}
        - a_{\ell j}
    }{
        b_{\ell j}-a_{\ell j}
    }.
    \label{eq:style_scaling}
\end{equation}
Features with coincident lower and upper thresholds were assigned a normalized value of zero for all repositories.

Let $\mathcal{R}_{\ell q}$ denote the eligible RQ2 repositories in language $\ell$ and creation quarter $q$, and let $n_{\ell q}=|\mathcal{R}_{\ell q}|$. For $n_{\ell q}\geq 2$, we calculated
\begin{equation}
    D_{\ell q}
    =
    \frac{2}{n_{\ell q}(n_{\ell q}-1)}
    \sum_{\substack{r,s\in\mathcal{R}_{\ell q}\\ r<s}}
    \left\|
        \mathbf{z}_{r}^{(2)}-\mathbf{z}_{s}^{(2)}
    \right\|_2,
    \label{eq:cohort_diversity}
\end{equation}
where $r<s$ indexes each unordered pair once. Each distinct repository pair received equal weight. Lower values indicate less dispersion in naming styles within a cohort.

We compared $D_{\ell q}$ across creation cohorts within each language. Because normalization was language-specific, our emphasis is on within-language patterns rather than absolute rankings of diversity across languages. This statistic measures dispersion within a cohort; it does not track changes in individual repositories or identify a particular naming style toward which repositories are converging. The relationship between creation time and the timing of the analyzed code is assessed in Section~\ref{sec:data_collection} and Appendix~\ref{app:repository_creation_and_observed_code}.

\subsection{Measuring Network Locality of Naming Styles}
\label{sec:network_locality}

To address RQ3, we related naming-style similarity to the shortest-path distance between repository owners in a collaboration network. We constructed an undirected, unweighted graph from GH Archive~\cite{gharchive} PushEvents recorded between January 1, 2010 and December 31, 2025. Accounts with login names ending in \texttt{[bot]} were excluded. Two distinct users were connected when both had pushed to the same repository. For graph construction, we retained only repositories with at most 20 distinct remaining pushing users over the network observation period, limiting the contribution of projects involving large numbers of users. The largest connected component contained 8{,}997{,}065 users and 29{,}657{,}442 links.

%
%

We used this fixed graph for all creation cohorts. Shortest paths could pass through any user in the LCC, including users who did not own a repository in the naming-style sample. Network distances therefore represent proximity in the cumulative collaboration graph rather than the collaboration network as it existed in each historical quarter.

For each language and creation quarter, we compared distinct owners of eligible repositories in the RQ3 sample. Naming features were normalized using the clipping and scaling procedure described in Section~\ref{sec:cohort_diversity}, with parameters fitted separately for each language using all eligible RQ3 repositories pooled across creation cohorts. These parameters were held fixed across quarters. Let Let
$d^{(3)}(r,s)=\|\mathbf{z}_{r}^{(3)}-\mathbf{z}_{s}^{(3)}\|_2$
denote the resulting naming-style distance between repositories $r$ and $s$.

To prevent owners with multiple repositories from contributing an unrestricted number of repository pairs to the aggregate, we first summarized distances at the owner-pair level. Within a given language and creation quarter, let $\mathcal{U}$ be the set of eligible owners and $\mathcal{R}_u$ the repositories owned by $u$. For distinct owners $u$ and $v$, we calculated
\begin{equation}
    \delta(u,v)
    =
    \frac{
        \displaystyle
        \sum_{r\in\mathcal{R}_u}
        \sum_{s\in\mathcal{R}_v}
        d^{(3)}(r,s)
    }{
        |\mathcal{R}_u|\,|\mathcal{R}_v|
    }.
    \label{eq:owner_pair_style_distance}
\end{equation}
Thus, each owner-pair distance averages all cross-owner repository combinations within the language and cohort. Repository ownership provides the link between a codebase and a network position; the measured style characterizes the repository's code rather than exclusively the owner's own contributions.

Next, for each focal owner $u$, we averaged owner-pair distances over eligible target owners at the same hop distance. Let $d_G(u,v)$ be the shortest-path distance in the fixed graph and define
$\mathcal{T}_h(u)=\{v\in\mathcal{U}\setminus\{u\}:d_G(u,v)=h\}$.
For nonempty $\mathcal{T}_h(u)$, the focal-owner mean was
\begin{equation}
    m_h(u)
    =
    \frac{1}{|\mathcal{T}_h(u)|}
    \sum_{v\in\mathcal{T}_h(u)}
    \delta(u,v).
    \label{eq:focal_owner_hop_average}
\end{equation}
Finally, let
$\mathcal{F}_h=\{u\in\mathcal{U}:|\mathcal{T}_h(u)|>0\}$
denote focal owners with at least one eligible target at distance $h$. The cohort-level mean at that distance was
\begin{equation}
    L_h
    =
    \frac{1}{|\mathcal{F}_h|}
    \sum_{u\in\mathcal{F}_h}m_h(u).
    \label{eq:network_locality}
\end{equation}
This procedure gives equal weight to target owners within each focal owner's hop group and equal weight to contributing focal owners, rather than pooling all repository pairs. The set of contributing focal owners can differ across hop distances and creation cohorts.

We examined hop-specific distances from 1 to 10. Figure~\ref{fig:network_and_naming_distance} displays 1--3 hops, while the full set of 1--10 hops is shown in Appendix~\ref{app:detailed_rq3_results}, Figure~\ref{fig:network_and_naming_distance_all_hops}. In both figures, hop-specific points with fewer than 100 contributing focal owners were omitted. The overall reference curve, labeled ``All pairs,'' represents the equally weighted mean of owner-pair distances $\delta(u,v)$ over all distinct owner pairs within each language and creation quarter, including pairs separated by more than 10 hops. The 100-owner display threshold does not affect this overall reference. Counts of contributing focal owners by language, creation quarter, and hop distance are reported in Appendix~\ref{app:sample_coverage}, Figure~\ref{fig:rq3_focal_owner_coverage}.

We assessed network locality by comparing naming-style distances across hop distances within each language and creation cohort, and examined whether the relationship persisted across cohorts. Because RQ2 and RQ3 use different samples and aggregation schemes, their overall distance curves need not coincide. These comparisons describe the association between naming-style similarity and cumulative collaboration-network proximity; they do not identify a causal effect of collaboration or establish that a convention propagated along a particular network path.

\section{Results}
\label{sec:results}

\subsection{RQ1: Naming Styles and LLM-Related Commits}
\label{sec:results_rq1}

\begin{figure}[tbp]
    \centering
    \includegraphics[width=\linewidth]{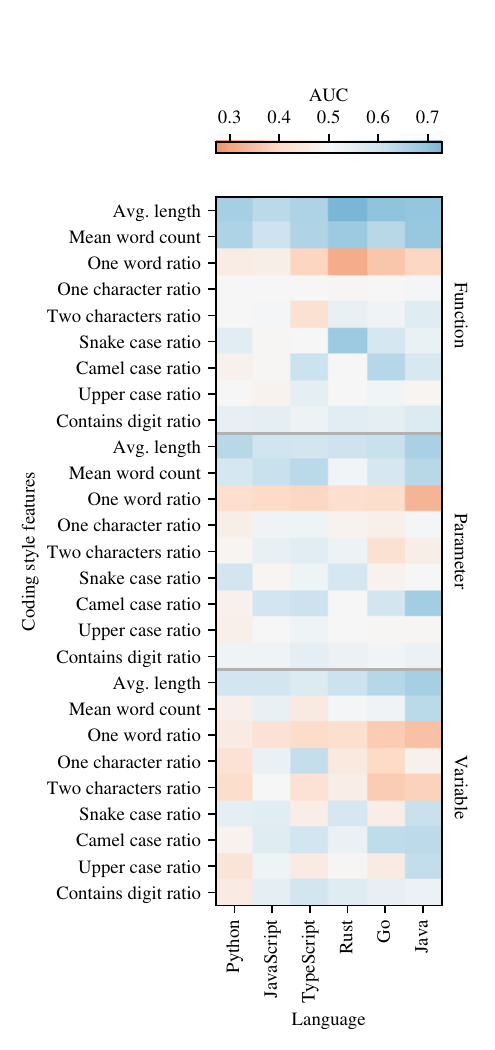}
    \caption{
        Differences in identifier naming features between repositories
        with and without detected LLM-related commits.
        Rows show the 27 features of function, parameter, and variable
        names; columns show six programming languages.
        Colors encode Mann--Whitney AUC: blue ($AUC>0.5$) indicates
        a tendency toward higher values in the detected group,
        red ($AUC<0.5$) indicates the opposite tendency,
        and white corresponds to $AUC=0.5$.
    }
    \label{fig:llm_related_naming_features}
\end{figure}

To address RQ1, we examine whether repositories with detected LLM-related commits exhibit different identifier naming characteristics from repositories without such detections. We compare the two groups among repositories created in September 2025 (Section~\ref{sec:rq1_comparison}). Figure~\ref{fig:llm_related_naming_features} summarizes the direction and magnitude of the differences across the 27 naming features, and Appendix~\ref{app:detailed_results}, Figure~\ref{fig:rq1_group_means}, reports the corresponding group means.

The most consistent difference is identifier length. Across all six programming languages, the detected group has longer identifiers on average for functions, parameters, and variables. For example, the mean function-name length in Python is 15.99 characters in the detected group compared with 13.69 in the not-detected group. The corresponding values are 16.5 versus 12.7 characters for Rust and 14.5 versus 11.8 for Java. Function and parameter names also contain more constituent words on average in the detected group in all six languages. In Python, for instance, function names contain 2.75 words on average in the detected group versus 2.45 in the not-detected group, while parameter names contain 1.98 versus 1.74 words. Differences in the number of words in variable names are smaller and less consistent across languages, even though their average character lengths remain higher in the detected group.

Naming-case differences are language-specific but exhibit another systematic pattern. For function names, the detected group shows greater use of the case pattern that is already prevalent in each language in our sample. The proportion represented by the \texttt{snake\_case} feature is higher for Python (80.2\% versus 73.2\%) and Rust (77.9\% versus 66.0\%). Conversely, the proportion represented by the \texttt{camelCase} feature is higher for JavaScript (84.0\% versus 81.0\%), TypeScript (74.4\% versus 60.9\%), Go (93.8\% versus 88.6\%), and Java (85.6\% versus 77.8\%). Similar directional differences are also visible for parameter and variable names in the detailed results in Appendix~\ref{app:detailed_results}. Thus, the detected group does not shift toward a single naming format shared across languages; rather, the differences tend to reinforce patterns that are already common within each language.

Taken together, RQ1 reveals systematic naming-style differences associated with detected LLM-related commits. Repositories in the detected group use longer identifiers, function and parameter names composed of more words, and, across languages, greater use of locally prevalent naming-case patterns. These results provide an empirical association between LLM-related development activity and naming conventions, while the language-specific direction of the case-pattern differences suggests reinforcement of existing conventions rather than convergence toward a universal naming style. Because the groups are defined by detected LLM-related commit messages, however, these comparisons do not establish that LLM assistance caused the observed differences.

\subsection{RQ2: Convergence across Repository Creation Cohorts}
\label{sec:results_rq2}

The group differences observed in RQ1 do not by themselves indicate whether naming conventions have become more or less diverse across the broader population of projects. To address RQ2, we therefore examine how within-language naming-style diversity varies across repository creation cohorts, using the within-cohort pairwise distances defined in Section~\ref{sec:cohort_diversity}.

\begin{figure}[tbp]
    \centering
    \includegraphics[width=0.93\linewidth]{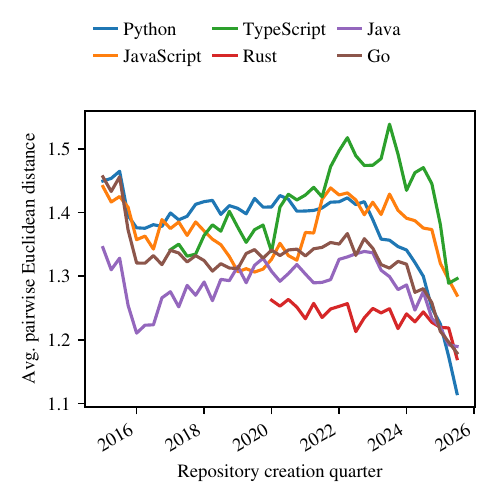}
    \caption{
        Identifier naming-style diversity across repository creation
        cohorts in six programming languages.
        Each point shows the mean pairwise Euclidean distance between
        repositories within the same language and creation quarter.
        Lower distances indicate less within-cohort dispersion.
        Features are normalized separately within each language as
        described in Section~\ref{sec:cohort_diversity}; absolute
        distance levels should therefore not be compared across
        languages.
    }
    \label{fig:cohort_naming_diversity}
\end{figure}

Figure~\ref{fig:cohort_naming_diversity} shows a broad decline in naming-style distance among recent creation cohorts, although the trajectories differ substantially across languages. The clearest pattern appears in Python. Naming-style distance is comparatively stable across much of the earlier period, but then declines markedly in recent cohorts: the mean pairwise distance decreases from 1.389 in 2023 Q1 to 1.115 in 2025 Q3, a reduction of 0.274, or approximately 19.7\% on this normalized distance scale. Thus, Python repositories created in recent cohorts are substantially less dispersed in their naming features than those created only a few years earlier.

The other languages also show lower distances in their most recent cohorts, but they reach these levels through different trajectories. JavaScript declines during the earlier period, rises again around the early 2020s, and then decreases in the latest cohorts. TypeScript exhibits an especially pronounced increase in diversity through its earlier cohorts before reversing direction and declining sharply toward the end of the observation period. Java and Go likewise show noticeable decreases in recent cohorts after earlier periods of relative stability or increase. Rust covers a shorter observation period and exhibits smaller fluctuations, but its latest cohorts also show lower distances than those immediately preceding them. The convergence is therefore broad across languages, but neither monotonic nor synchronized.

Taken together, these patterns answer RQ2 by showing broad but temporally heterogeneous convergence across repository creation cohorts. Recent cohorts tend to exhibit lower within-language naming-style diversity, but the timing and path of this convergence differ across programming languages. Importantly, ``convergence'' here refers to a reduction in dispersion among repositories: the analysis does not show that repositories are moving toward a single identifiable naming style, nor that they are converging specifically toward the feature profile observed in the detected group in RQ1.

The strongest declines occur during the period in which LLM-based development tools became widely used~\cite{vibecoding_survey}, making the temporal correspondence relevant to the broader question motivating this study. However, the cohort comparison alone does not identify LLM diffusion as the cause of the decline. Moreover, because repositories are grouped by creation date while their source code was collected retrospectively, the temporal interpretation relies on the approximate correspondence between repository creation and observed code timing described in Section~\ref{sec:data_collection} and Appendix~\ref{app:repository_creation_and_observed_code}.

\subsection{RQ3: Network Locality amid Overall Convergence}
\label{sec:results_rq3}

RQ2 shows that repositories in recent creation cohorts tend to be more similar in naming style overall. RQ3 asks whether this broader convergence is accompanied by the disappearance of differences associated with collaboration-network proximity, or whether network-local variation remains visible. We therefore compare naming-style distances across repository-owner pairs separated by different shortest-path distances in the collaboration network (Section~\ref{sec:network_locality}).

\begin{figure*}[tbp]
    \centering
    \includegraphics[width=1\linewidth]{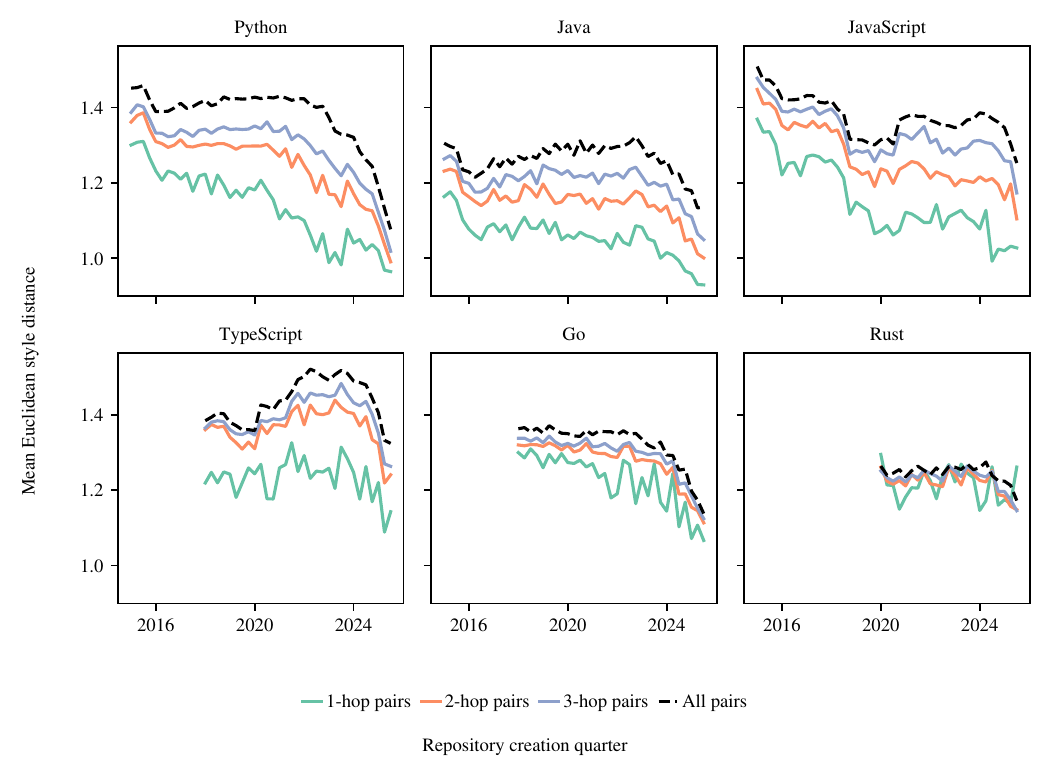}
    \caption{
        Naming-style distance by repository creation quarter and shortest-path separation between repository owners in the fixed collaboration graph.
        Panels show six programming languages.
        Colored curves represent separations of 1, 2, and 3 hops;
        black dashed curves show the equally weighted mean over all distinct owner pairs, regardless of hop distance.
        Hop-specific distances use the owner-balanced aggregation defined in Section~\ref{sec:network_locality}.
        Points with fewer than 100 contributing focal owners are omitted.
        Results for 1--10 hops are shown in Figure~\ref{fig:network_and_naming_distance_all_hops}.
    }
    \label{fig:network_and_naming_distance}
\end{figure*}

Figure~\ref{fig:network_and_naming_distance} shows a clear and recurring relationship between network proximity and naming-style similarity in most languages. For Python, JavaScript, TypeScript, Java, and Go, owner pairs separated by one hop generally exhibit smaller naming-style distances than those separated by two hops, which in turn tend to be smaller than those for three-hop pairs. This ordering persists across substantial portions of the observation period, although it is not perfectly monotonic in every quarter. Rust differs from the other languages: its hop-specific curves overlap and fluctuate, and a comparable ordering is not consistently visible. Thus, network locality is evident in five of the six languages studied, rather than being a universal pattern.

The central result is that this proximity-related ordering remains visible even as overall naming-style distances decline in recent cohorts. In Python and Java, for example, the 1-hop, 2-hop, and 3-hop curves move downward along with the overall reference curve while remaining distinctly ordered. Similar persistence is visible in JavaScript, TypeScript, and Go over much of their recent periods. Broad convergence therefore reduces absolute differences among repositories without making network proximity irrelevant: repositories owned by socially closer users continue to be more similar than those owned by more distant users.

The dashed ``All pairs'' curve provides an internal reference for the overall level of naming-style distance within the RQ3 sample. Because it is calculated using a different sample and aggregation scheme from the RQ2 statistic, its absolute values should not be directly compared with those in Figure~\ref{fig:cohort_naming_diversity}. Its role here is to show that the hop-specific ordering persists while the overall level of distance among repository owners changes.

The magnitude of locality is not constant over time. In Python, Java, and TypeScript, the figure suggests that the absolute gaps between nearby and more distant owners widen during some earlier cohorts and narrow again in more recent ones as overall distances decrease. This secondary pattern should be distinguished from the more robust finding that the ordering itself persists. A narrowing absolute gap does not by itself imply that network locality has weakened relative to overall naming-style diversity.

Taken together, RQ3 shows that broad convergence and network-local variation coexist. In most of the studied languages, recent cohorts exhibit lower overall naming-style distances while preserving systematic differences associated with collaboration-network proximity. The finding therefore distinguishes two aspects of convention structure: how much variation exists across repositories and how the remaining variation is organized across social relationships. Greater population-level similarity does not necessarily imply that conventions become independent of network structure.

\section{Discussion}
\label{sec:discussion}

\subsection{Implications}
\label{sec:implications}

Our central finding is that broad convergence and network locality coexist. In most of the studied languages, recent repository creation cohorts exhibit lower naming-style dispersion, yet repositories whose owners are closer in the collaboration network remain more similar. This result connects to a longstanding question in the study of social conventions: how can widely shared practices emerge while socially structured differences persist? Prior work has shown that local coordination can generate both shared conventions and persistent differentiation depending on network structure~\cite{Axelrod_local_convention_global_pol,Centola_Baronchelli_cordination_structure}. Our findings highlight that these are two distinct properties of convention structure: the overall amount of variation and the way that variation is organized across social relationships. A population can therefore become more homogeneous without becoming socially unstructured. This distinction is important for studying cultural change in environments where new population-scale sources of coordination coexist with established social networks.

The results also suggest a useful way to think about the role of LLMs in such environments. From the perspective of machine culture, intelligent systems can participate in the transmission and selection of cultural practices~\cite{brinkmann2023machine_culture}. LLMs provide a widely shared source of suggestions that is accessible to developers who may be distant from one another in collaboration networks, while local project requirements and interactions continue to shape how those suggestions are incorporated. Our RQ1 results are consistent with this layered form of coordination: repositories with detected LLM-related commits tend to use longer identifiers and, in several languages, make greater use of naming patterns already prevalent in that language, rather than moving toward a single universal naming format. Thus, AI-related cultural change may involve not only the introduction of new practices, but also the amplification of existing conventions in context-dependent ways.

This coexistence of convergence and locality also has practical implications. Greater consistency in routine naming choices may reduce coordination costs and make codebases easier to navigate across developers~\cite{Allamanis_2014}, while excessive homogenization could reduce exposure to alternative practices. Similar tensions between individual benefits and collective diversity have been observed in other forms of LLM-assisted production~\cite{Doshi_Hauser_diverisity}. Our results suggest that coding assistants should therefore not be evaluated only by how well they reproduce broadly common conventions, but also by whether they can accommodate project- and community-specific practices. More generally, the goal need not be maximum uniformity, but rather reducing unnecessary coordination costs while preserving locally meaningful variation.

\subsection{Limitations}
\label{sec:limitations}

Several limitations qualify the scope of our findings. First, our indicator of LLM-related activity is a proxy rather than a direct measure of LLM use in producing the analyzed code. A matching commit message may refer to LLM-related functionality rather than coding assistance, whereas LLM assistance may occur without being mentioned in a commit message. The detected and not-detected groups may therefore differ for reasons other than LLM use. In addition, RQ1 identifies an association between detected LLM-related activity and naming style, while the temporal correspondence observed in RQ2 and RQ3 does not isolate LLM diffusion from other changes in software-development practices. The three analyses should therefore be interpreted as complementary observational evidence rather than as a causal chain showing that LLM use produced the observed convergence.

Second, temporal attribution is approximate. We analyze the default-branch HEAD obtained at clone time and organize repositories by their creation dates rather than reconstructing source-code snapshots at the time of creation. Among audited repositories created before 2025 with usable timestamps, repository creation and HEAD-commit author dates fall in the same calendar quarter for 68.0\% of the RQ2 sample and 69.9\% of the RQ3 sample. These results indicate substantial but incomplete temporal alignment. Later modifications can therefore be attributed to earlier creation cohorts, and differences across cohorts may partly reflect project age or maturity. Accordingly, our temporal results are best interpreted as retrospective comparisons across repository creation cohorts rather than direct longitudinal observations of the same projects over time.

Third, the network results characterize associations with cumulative collaboration proximity rather than pathways of social influence~\cite{Aral_dintinguishing_contagion_homophily}. The fixed collaboration graph aggregates relationships over the full observation period, so distances for earlier creation cohorts may include collaborations established later. Similar naming styles among nearby owners may also arise from homophily~\cite{McPherson_homophily} or selection into collaboration, shared project requirements, or common code, rather than interpersonal transmission alone. Moreover, repository ownership is not equivalent to code authorship: the naming features summarize the repository's code, which may contain contributions from multiple developers, whereas the owner is used to locate that repository in the collaboration network. RQ3 should therefore be interpreted as evidence that naming-style variation is structured by cumulative collaboration proximity, not that conventions necessarily diffused along the observed network paths.

Finally, the sampling and measurement design limits generalizability. Our data cover public, non-fork GitHub repositories in six programming languages and within the specified repository-size range; RQ3 further focuses on owners in the largest connected component of the collaboration graph. Eligibility for the final analysis also requires extractable function, parameter, and variable names, which can alter the composition of the analyzed samples across languages and groups. The results therefore need not generalize to private repositories, very large projects, or software development outside GitHub. In addition, identifier naming captures only one dimension of coding convention. Our features describe formal properties of names, not their semantic appropriateness, readability, or software quality. Likewise, lower dispersion indicates greater similarity across repositories but does not identify a unique style toward which they converge, and persistent hop-distance ordering does not imply that the relative strength of network locality remains constant over time.

\section{Conclusion}

We examined how identifier naming conventions are distributed across GitHub repositories in the context of the rapid diffusion of LLM-based development tools. Across six programming languages, repositories with detected LLM-related commits exhibited systematically different naming characteristics, including longer identifiers and, in several languages, greater use of naming patterns already prevalent in that language. We also found that naming-style diversity is lower among recent repository creation cohorts, although the timing and trajectory of this convergence vary across languages. Most importantly, this broader convergence does not eliminate network-local variation: in five of the six languages, repositories whose owners are closer in the collaboration network remain more similar in naming style even in recent, more homogeneous cohorts. These findings show that aggregate convergence and network locality can coexist. More broadly, understanding how widely shared AI systems reshape social conventions requires examining not only whether practices become more similar overall, but also how the remaining variation continues to be structured by human social relationships.

\section*{Acknowledgments}
\textbf{Funding.}
This work was supported in part by The Nippon Foundation HUMAI Program, JST SPRING Grant No. JPMJSP2124, JSPS KAKENHI Grant No. JP25K03105, and JST ERATO Grant No. JPMJER2502.

\noindent\textbf{Generative AI Disclosure.}
ChatGPT and OpenAI Codex were used to assist with proofreading and code generation. All generated outputs were carefully reviewed and verified by the authors. The authors take full responsibility for all content, analyses, and results presented in this paper.


\bibliography{aaai25}

@article{brinkmann2023machine_culture,
  title={Machine culture},
  author={Brinkmann, Levin and Baumann, Fabian and Bonnefon, Jean-Fran{\c{c}}ois and Derex, Maxime and M{\"u}ller, Thomas F and Nussberger, Anne-Marie and Czaplicka, Agnieszka and Acerbi, Alberto and Griffiths, Thomas L and Henrich, Joseph and others},
  journal={Nature Human Behaviour},
  volume={7},
  number={11},
  pages={1855--1868},
  year={2023},
}

@inproceedings{Kooti_Emergence_of_osn2012,
  title     = {The Emergence of Conventions in Online Social Networks},
  author    = {Kooti, Farshad and Yang, Haeryun and Cha, Meeyoung and Gummadi, Krishna P. and Mason, Winter A.},
  booktitle = {Proceedings of the International AAAI Conference on Web and Social Media},
  pages     = {194--201},
  volume={6},
  year      = {2012},
  doi       = {10.1609/icwsm.v6i1.14267}
}

@inproceedings{Rotabi_2017,
 title={Tracing the Use of Practices Through Networks of Collaboration}, volume={11}, ISSN={2162-3449},doi={10.1609/icwsm.v11i1.14870}, number={1}, booktitle={Proceedings of the International AAAI Conference on Web and Social Media}, author={Rotabi, Rahmtin and Danescu-Niculescu-Mizil, Cristian and Kleinberg, Jon}, year={2017}, pages={201–209} }

@article{Centola_Baronchelli_cordination_structure,
author = {Damon Centola  and Andrea Baronchelli },
title = {The spontaneous emergence of conventions: An experimental study of cultural evolution},
journal = {Proceedings of the National Academy of Sciences},
volume = {112},
number = {7},
pages = {1989--1994},
year = {2015},
doi = {10.1073/pnas.1418838112},
}

@article{Kodak_delving,

author = {Kobak, Dmitry and
          Gonz{\'a}lez-M{\'a}rquez, Rita and
          Horv{\'a}t, Em{\H{o}}ke-{\'A}gnes and
          Lause, Jan},
title = {Delving into LLM-assisted writing in biomedical publications through excess vocabulary},
journal = {Science Advances},
volume = {11},
number = {27},
pages = {eadt3813},
year = {2025},
doi = {10.1126/sciadv.adt3813},
}

@article{
Doshi_Hauser_diverisity,
author = {Anil R. Doshi  and Oliver P. Hauser },
title = {Generative AI enhances individual creativity but reduces the collective diversity of novel content},
journal = {Science Advances},
volume = {10},
number = {28},
pages = {eadn5290},
year = {2024},
doi = {10.1126/sciadv.adn5290},
}

@article{Acerbi_Stubbersfield,
author = {Alberto Acerbi  and Joseph M. Stubbersfield },
title = {Large language models show human-like content biases in transmission chain experiments},
journal = {Proceedings of the National Academy of Sciences},
volume = {120},
number = {44},
pages = {e2313790120},
year = {2023},
doi = {10.1073/pnas.2313790120},
}

@inproceedings{Allamanis_2014,
author = {Allamanis, Miltiadis and Barr, Earl T. and Bird, Christian and Sutton, Charles},
title = {Learning natural coding conventions},
year = {2014},
doi = {10.1145/2635868.2635883},
booktitle = {Proceedings of the 22nd ACM SIGSOFT International Symposium on Foundations of Software Engineering},
pages = {281-–293},
numpages = {13},
}

@INPROCEEDINGS{Butler_2011,
  author={Butler, Simon and Wermelinger, Michel and Yu, Yijun and Sharp, Helen},
  booktitle={2011 27th IEEE International Conference on Software Maintenance (ICSM)}, 
  title={Mining {Java} class naming conventions}, 
  year={2011},
  pages={93--102},
  doi={10.1109/ICSM.2011.6080776}}

@ARTICLE{Feitelson_2022,
  author={Feitelson, Dror G. and Mizrahi, Ayelet and Noy, Nofar and Shabat, Aviad Ben and Eliyahu, Or and Sheffer, Roy},
  journal={IEEE Transactions on Software Engineering}, 
  title={How Developers Choose Names}, 
  year={2022},
  volume={48},
  number={1},
  pages={37--52},
  doi={10.1109/TSE.2020.2976920}}

@INPROCEEDINGS{Zhang_2021,
  author={Zhang, Jingxuan and Zou, Weiqin and Huang, Zhiqiu},
  booktitle={2021 28th Asia-Pacific Software Engineering Conference (APSEC)}, 
  title={An Empirical Study on the Usage and Evolution of Identifier Styles in Practice}, 
  year={2021},
  pages={171--180},
  doi={10.1109/APSEC53868.2021.00025}}

@inproceedings{AlMadi2022Copilot,
author = {Al Madi, Naser},
title = {How Readable is Model-generated Code? Examining Readability and Visual Inspection of {GitHub Copilot}},
year = {2023},
doi = {10.1145/3551349.3560438},
booktitle = {Proceedings of the 37th IEEE/ACM International Conference on Automated Software Engineering},
pages={205:1--205:5}
}

@article{Liu2024ChatGPTCodeQuality,
author = {Liu, Yue and Le-Cong, Thanh and Widyasari, Ratnadira and Tantithamthavorn, Chakkrit and Li, Li and Le, Xuan-Bach D. and Lo, David},
title = {Refining ChatGPT-Generated Code: Characterizing and Mitigating Code Quality Issues},
year = {2024},
volume = {33},
number = {5},
doi = {10.1145/3643674},
journal = {ACM Transactions on Software Engineering and Methodology},
pages={116:1--116:26}
}

@article{Wang2025CodingStyle,
author = {Wang, Yanlin and Jiang, Tianyue and Liu, Mingwei and Chen, Jiachi and Mao, Mingzhi and Liu, Xilin and Ma, Yuchi and Zheng, Zibin},
title = {Beyond Functional Correctness: Investigating Coding Style Inconsistencies in Large Language Models},
year = {2025},
volume = {2},
number = {FSE},
doi = {10.1145/3715749},
journal = {Proceedings of the ACM on Software Engineering},
pages={690--712},
}

@article{Young_evoluoin_of_convention,
 author = {H. Peyton Young},
 journal = {Econometrica},
 number = {1},
 pages = {57--84},
 title = {The Evolution of Conventions},
 volume = {61},
 year = {1993}
}

@article{McPherson_homophily,
 author = {Miller McPherson and Lynn Smith-Lovin and James M. Cook},
 journal = {Annual Review of Sociology},
 pages = {415--444},
 publisher = {Annual Reviews},
 title = {Birds of a Feather: Homophily in Social Networks},
 volume = {27},
 year = {2001}
}

@inproceedings{xu-etal-2026-code,
    title = "code-transformed: The Influence of Large Language Models on Code",
    author = "Xu, Yuliang  and
      Huang, Siming  and
      Geng, Mingmeng  and
      Wan, Yao  and
      Shi, Xuanhua  and
      Chen, Dongping",
    booktitle = "Findings of the {A}ssociation for {C}omputational {L}inguistics: {EACL} 2026",
    year = "2026",
    doi = "10.18653/v1/2026.findings-eacl.290",
    pages = "5462--5490",
}

@INPROCEEDINGS{Lawrie_naming_style_length,
  author={Lawrie, D. and Morrell, C. and Feild, H. and Binkley, D.},
  booktitle={14th IEEE International Conference on Program Comprehension (ICPC'06)}, 
  title={What's in a Name? A Study of Identifiers}, 
  year={2006},
  pages={3--12},
  doi={10.1109/ICPC.2006.51}}

@article{Binkley_identifier_name,
title = {Identifier length and limited programmer memory},
journal = {Science of Computer Programming},
volume = {74},
number = {7},
pages = {430--445},
year = {2009},
issn = {0167-6423},
doi = {https://doi.org/10.1016/j.scico.2009.02.006},
author = {Dave Binkley and Dawn Lawrie and Steve Maex and Christopher Morrell},
}

@INPROCEEDINGS{Binkley_camel_underscore,
  author={Binkley, David W. and Davis, Marcia and Lawrie, Dawn J. and Morrell, Christopher},
  booktitle={2019 IEEE/ACM 27th International Conference on Program Comprehension (ICPC)}, 
  title={To {CamelCase} or under\_score}, 
  year={2019},
  pages={177},
  doi={10.1109/ICPC.2019.00035}}

@article{Hofmeister_2018, title={Shorter identifier names take longer to comprehend}, volume={24},doi={10.1007/s10664-018-9621-x}, number={1}, journal={Empirical Software Engineering}, author={Hofmeister, Johannes C. and Siegmund, Janet and Holt, Daniel V.}, year={2019}, pages={417--443} }

@article{Axelrod_local_convention_global_pol,
 author = {Robert Axelrod},
 journal = {The Journal of Conflict Resolution},
 number = {2},
 pages = {203--226},
 title = {The Dissemination of Culture: A Model with Local Convergence and Global Polarization},
 volume = {41},
 year = {1997}
}

@article{Centola_long_tie,
 author = {Damon Centola and Michael Macy},
 journal = {American Journal of Sociology},
 number = {3},
 pages = {702--734},
 title = {Complex Contagions and the Weakness of Long Ties},
 volume = {113},
 year = {2007}
}

@inproceedings{Lima_coding_togther_at_scale,
  author    = {Lima, Antonio and Rossi, Luca and Musolesi, Mirco},
  title     = {Coding Together at Scale: {GitHub} as a Collaborative Social Network},
  booktitle = {Proceedings of the International AAAI Conference on Web and Social Media},
  volume    = {8},
  number    = {1},
  pages     = {295--304},
  year      = {2014},
  doi       = {10.1609/icwsm.v8i1.14552},
}

@inproceedings{community_toxic_norm,
  title     = "Quick, community-specific learning: How distinctive toxicity
               norms are maintained in political subreddits",
  author    = "Rajadesingan, Ashwin and Resnick, Paul and Budak, Ceren",
  booktitle   = "Proceedings of the International AAAI Conference on Web and
               Social Media",
  volume    =  14,
  pages     = "557--568",
  year      =  2020,
  doi = {10.1609/icwsm.v14i1.7323}
}

@inproceedings{Khalid_2020_style_community, title={Style Matters! Investigating Linguistic Style in Online Communities}, volume={14}, doi={10.1609/icwsm.v14i1.7306}, booktitle={Proceedings of the International AAAI Conference on Web and Social Media}, author={Khalid, Osama and Srinivasan, Padmini}, year={2020}, pages={360--369} }

@article{
Aral_dintinguishing_contagion_homophily,
author = {Sinan Aral  and Lev Muchnik  and Arun Sundararajan },
title = {Distinguishing influence-based contagion from homophily-driven diffusion in dynamic networks},
journal = {Proceedings of the National Academy of Sciences},
volume = {106},
number = {51},
pages = {21544--21549},
year = {2009},
doi = {10.1073/pnas.0908800106},
}

@inproceedings{wikipedia_norms, title={Seeing Like an AI: How LLMs Apply (and Misapply) Wikipedia Neutrality Norms}, volume={20},doi={10.1609/icwsm.v20i1.42630}, number={1}, booktitle={Proceedings of the International AAAI Conference on Web and Social Media}, author={Ashkinaze, Joshua and Guan, Ruijia and Kurek, Laura and Adar, Eytan and Budak, Ceren and Gilbert, Eric}, year={2026}, pages={146--173} }

@article{vibecoding_survey,
title = {Using AI-based coding assistants in practice: State of affairs, perceptions, and ways forward},
journal = {Information and Software Technology},
volume = {178},
pages = {107610},
year = {2025},
doi = {0.1016/j.infsof.2024.107610},
author = {Agnia Sergeyuk and Yaroslav Golubev and Timofey Bryksin and Iftekhar Ahmed},
}

@misc{gharchive,
  author = {Grigorik, Ilya},
  title = {GH Archive: Recording the public GitHub timeline},
  howpublished = {\url{https://www.gharchive.org/}},
  year = {2011}
}

\clearpage

\section*{Appendix}
\appendix

\section{LLM-Related Commit Search Records}
\label{app:llm_commit_records}

\begin{figure}[tbp]
    \centering
    \includegraphics[width=\linewidth]{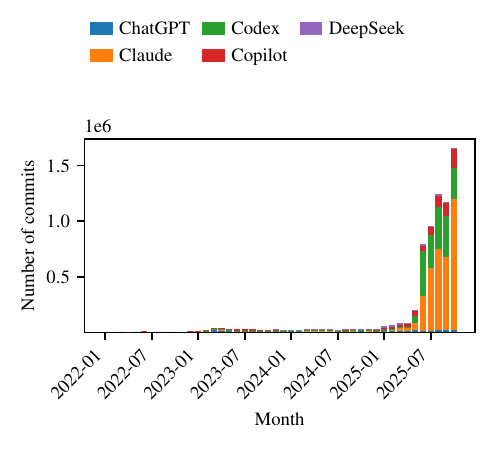}
    \caption{
        Monthly counts of LLM-related commit records from January 2022
        to October 2025, based on commit author timestamps in UTC.
        Stacked bars distinguish ChatGPT, Claude, Codex, Copilot, and
        DeepSeek. Records were counted separately in each keyword-specific
        dataset without deduplication within or across datasets; the counts
        therefore describe retrieved search records rather than unique
        commits or the prevalence of LLM adoption.
    }
    \label{fig:commitmessage_stats}
\end{figure}

Figure~\ref{fig:commitmessage_stats} summarizes the LLM-related commit
records retrieved using the keyword searches described in
Section~\ref{sec:llm_detection}.

\section{Temporal Alignment of Repository Creation and Observed Code}
\label{app:repository_creation_and_observed_code}
\begin{figure*}
    \centering
    \includegraphics[width=1\linewidth]{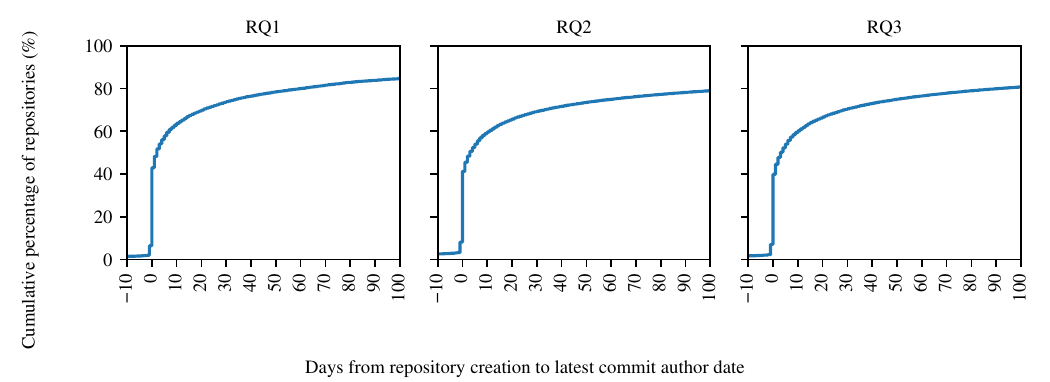}
    \caption{Cumulative distributions of elapsed completed days from repository creation to the saved HEAD commit's author timestamp. Panels show available clones for RQ1, RQ2, and RQ3, respectively, before final feature-eligibility filtering. RQ1 includes repositories created in September 2025, pooling both detection groups; RQ2 and RQ3 include repositories created before January 2025. The value at $-10$ days includes all earlier values, and intervals exceeding 100 days remain in the denominator.}
    \label{fig:repository_lifetime_cdf}
\end{figure*}

Figure~\ref{fig:repository_lifetime_cdf} shows the cumulative distributions of elapsed time between repository creation and the author timestamp of the saved HEAD commit. At day 30, the cumulative proportions are 73.9\% for RQ1, 69.4\% for RQ2, and 70.6\% for RQ3, including negative intervals. RQ1 includes repositories created in September 2025, whereas RQ2 and RQ3 include repositories created before January 2025. Negative intervals do not necessarily indicate errors: a newly created GitHub repository can contain an existing Git history with commits authored before the repository was created. However, Git author timestamps can also be set arbitrarily when commits are created or history is rewritten, so they may not reflect the actual last-modification times of the analyzed source files. These distributions therefore provide an approximate assessment of temporal alignment rather than evidence of historical code snapshots.
  
\section{Sample Coverage}
\label{app:sample_coverage}
Table~\ref{tab:repository-selection-flow} summarizes repository counts at each stage of the sampling and processing pipeline, separately by language and analysis sample. The final column reports repositories with extracted identifiers in all three categories: functions, parameters, and variables.

\begin{figure*}
    \centering
    \includegraphics[width=1\linewidth]{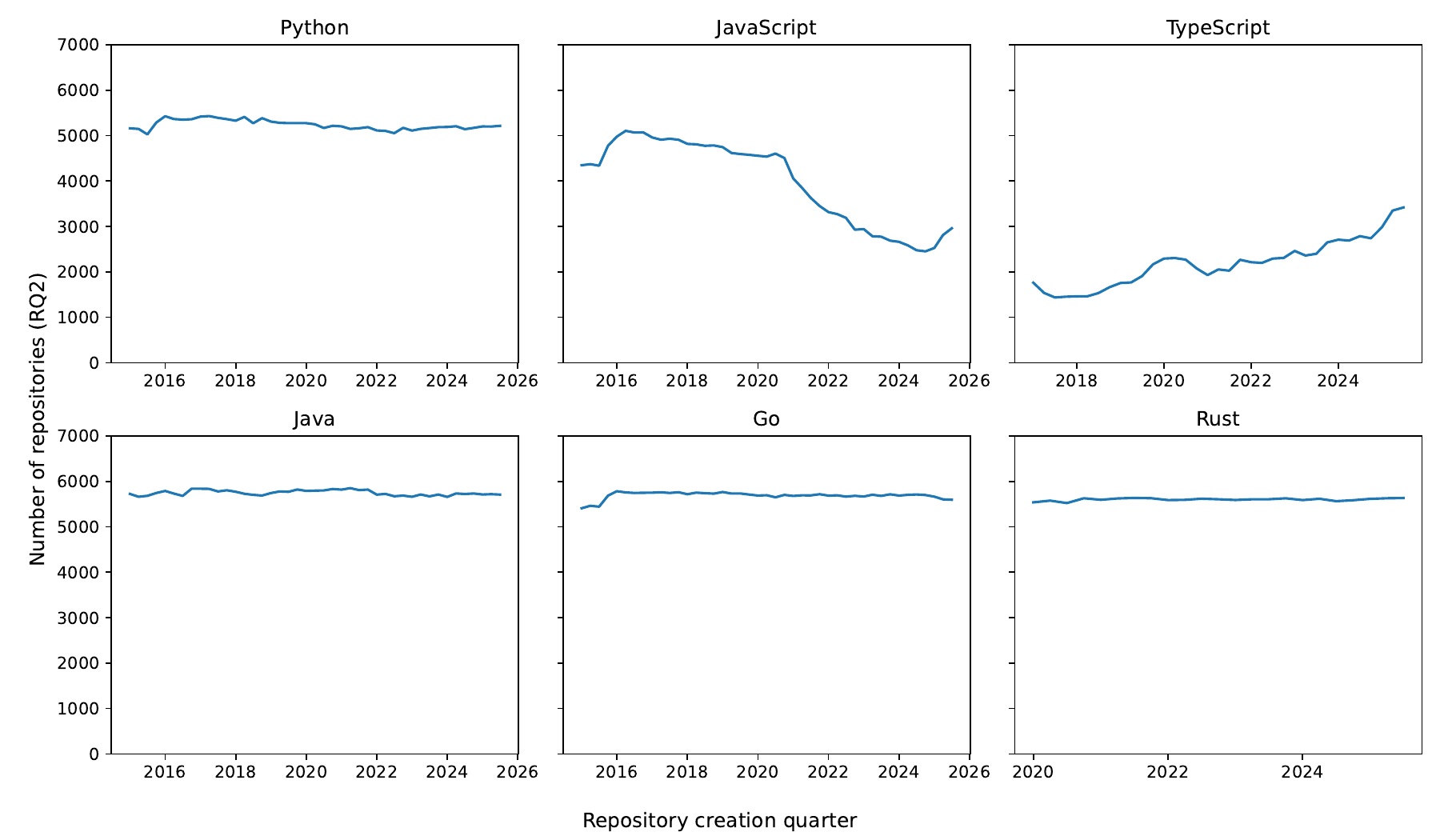}
    \caption{Numbers of repositories retained for RQ2 by repository creation quarter. Panels show the six programming languages. Counts include repositories with extracted function, parameter, and variable names that satisfy the final analysis criteria.}
    \label{fig:rq2_repository_coverage}
\end{figure*}

\begin{figure*}
    \centering
    \includegraphics[width=1\linewidth]{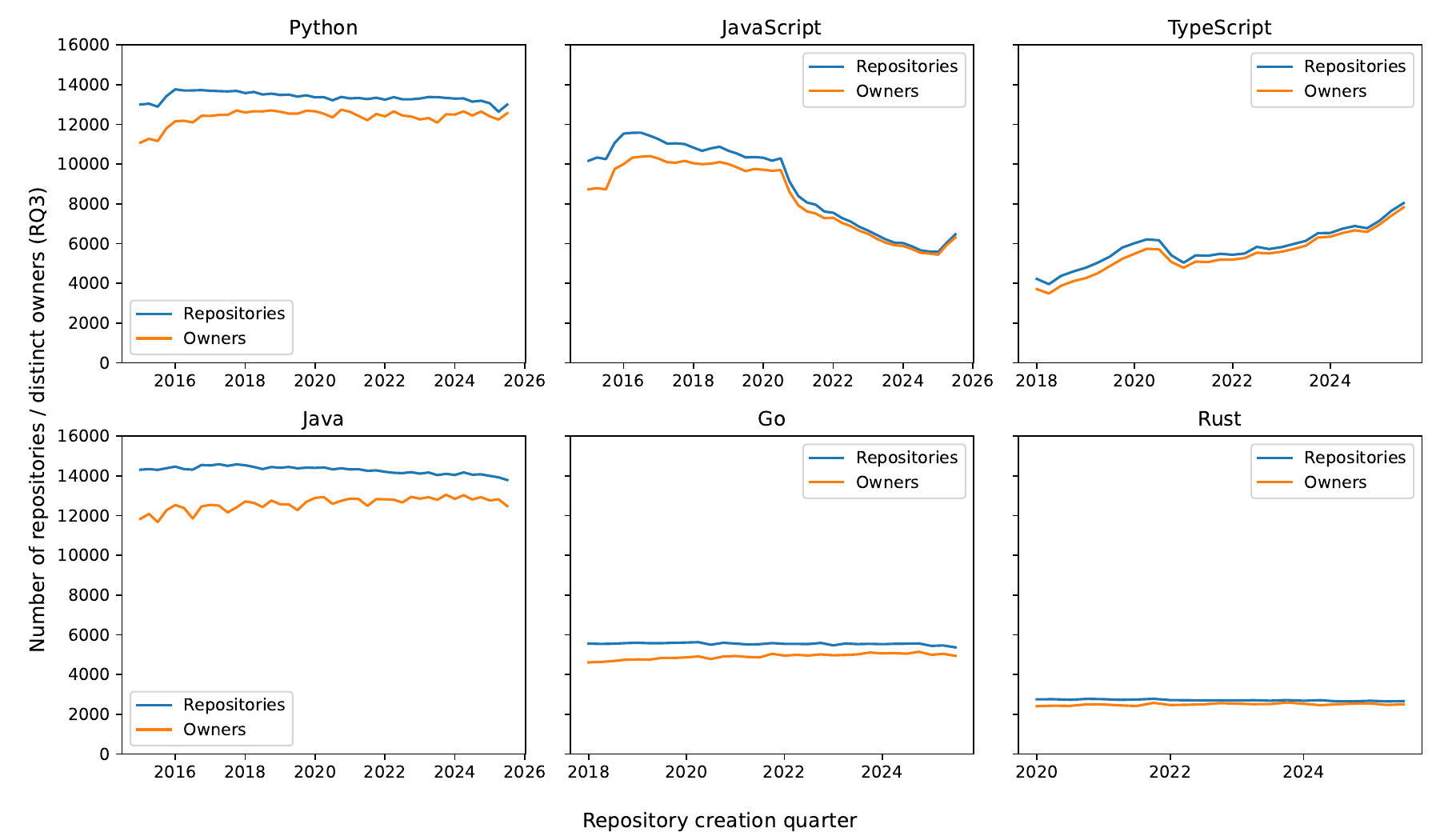}
    \caption{Numbers of repositories and distinct owners retained for RQ3 by repository creation quarter. Panels show the six programming languages. Owners are counted once within each language and quarter, regardless of how many eligible repositories they own. The same owner may contribute to multiple quarters or languages.}
    \label{fig:rq3_repository_owner_coverage}
\end{figure*}

Figure~\ref{fig:rq2_repository_coverage} shows the numbers of repositories retained for RQ2 by language and creation quarter. Figure~\ref{fig:rq3_repository_owner_coverage} shows the corresponding repository and distinct-owner counts for RQ3. The final RQ3 sample contains 1{,}237{,}319 distinct owners across all languages and creation quarters.

\begin{figure*}
    \centering
    \includegraphics[width=1\linewidth]{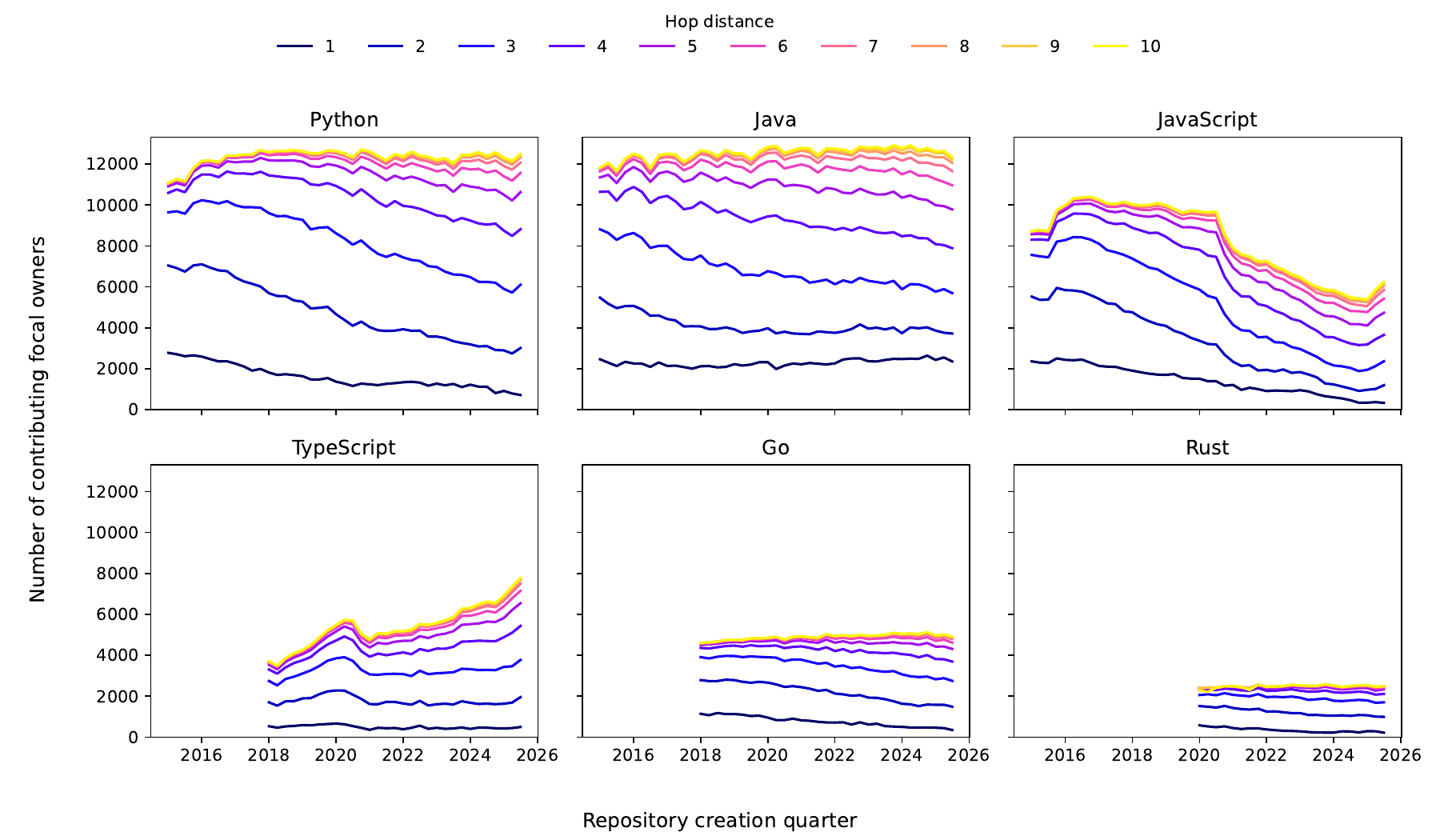}
    \caption{Contributing focal-owner counts by repository creation quarter and hop distance in RQ3. Panels show six programming languages, and colored lines represent distances of 1--10 hops, using the same colors as the main network-locality figure. Each owner is counted once per hop distance if at least one eligible target owner exists at that distance within the same language and creation quarter. Counts below the 100-owner display threshold used in the main figure are retained. All panels share the same vertical scale.}    \label{fig:rq3_focal_owner_coverage}
\end{figure*}

Figure~\ref{fig:rq3_focal_owner_coverage} reports the numbers of contributing focal owners at each hop distance from 1 to 10, separately by language and creation quarter. An owner contributes at distance $h$ if at least one other eligible owner in the same language and creation quarter is exactly $h$ hops away in the fixed collaboration graph. Counts below 100 are included here, although the corresponding naming-style distance estimates are omitted from the main figure.

\begin{table*}[tbp]
\centering
\small
\caption{Repository counts through the sampling, cloning, and feature-extraction pipeline.}
\label{tab:repository-selection-flow}
\begin{tabular}{lrrrr}
\toprule
Language & Candidate & Sampled & Clone succeeded & Final eligible repositories \\
\midrule
\multicolumn{5}{l}{\textit{RQ1: Detected}} \\
Python & 13,678 & 2,000 & 1,680 & 1,644 \\
JavaScript & 7,734 & 2,000 & 1,672 & 1,342 \\
TypeScript & 15,324 & 2,000 & 1,529 & 1,206 \\
Java & 1,200 & 1,200 & 1,055 & 1,032 \\
Go & 860 & 860 & 753 & 734 \\
Rust & 848 & 848 & 762 & 745 \\
\midrule
\multicolumn{5}{l}{\textit{RQ1: Not detected}} \\
Python & 327,563 & 2,000 & 1,764 & 1,572 \\
JavaScript & 454,061 & 2,000 & 1,779 & 828 \\
TypeScript & 374,224 & 2,000 & 1,712 & 957 \\
Java & 149,293 & 2,000 & 1,836 & 1,773 \\
Go & 18,016 & 2,000 & 1,762 & 1,700 \\
Rust & 12,985 & 2,000 & 1,821 & 1,713 \\
\midrule
\multicolumn{5}{l}{\textit{RQ2: Cohort}} \\
Python & 9,663,525 & 253,577 & 253,441 & 225,276 \\
JavaScript & 22,056,934 & 257,999 & 257,708 & 171,067 \\
TypeScript & 8,762,042 & 210,000 & 209,608 & 76,688 \\
Java & 11,649,264 & 258,000 & 257,084 & 247,191 \\
Go & 1,102,773 & 258,000 & 256,273 & 244,642 \\
Rust & 451,411 & 138,000 & 137,982 & 128,832 \\
\midrule
\multicolumn{5}{l}{\textit{RQ3: Network}} \\
Python & 3,334,770 & 645,000 & 632,305 & 574,575 \\
JavaScript & 7,943,176 & 645,000 & 632,172 & 384,484 \\
TypeScript & 3,424,390 & 465,000 & 454,662 & 179,876 \\
Java & 3,883,983 & 645,000 & 637,932 & 614,245 \\
Go & 443,034 & 186,000 & 182,300 & 171,738 \\
Rust & 258,466 & 69,000 & 67,411 & 62,226 \\
\bottomrule
\end{tabular}
\par\smallskip
\end{table*}

\section{Directory Exclusions}
\label{app:directory_exclusions}

Table~\ref{tab:directory_exclusions} lists the excluded directory names. Matching was exact, and all descendants of a matching directory were excluded. Each language-specific list was applied in addition to the common exclusions.

  \begin{table*}[tbp]
  \centering
  \small
  \caption{Directory-name exclusions used during identifier extraction.}
  \label{tab:directory_exclusions}
  \begin{tabular}{lp{0.65\textwidth}}
  \toprule
  Scope & Excluded directory names \\
  \midrule
  All languages &
  \texttt{.git}, \texttt{.github}, \texttt{coverage},
  \texttt{vendor}, \texttt{\_\_generated\_\_},
  \texttt{generated}, \texttt{gen} \\
  \midrule
  Python (additional) &
  \texttt{\_\_pycache\_\_}, \texttt{.venv}, \texttt{venv},
  \texttt{env}, \texttt{.tox}, \texttt{.mypy\_cache},
  \texttt{.pytest\_cache}, \texttt{dist}, \texttt{build},
  \texttt{site-packages} \\
  JavaScript / TypeScript (additional) &
  \texttt{node\_modules}, \texttt{dist}, \texttt{build},
  \texttt{out}, \texttt{.next}, \texttt{.nuxt},
  \texttt{.svelte-kit}, \texttt{.turbo},
  \texttt{.parcel-cache}, \texttt{.cache} \\
  Java (additional) &
  \texttt{target}, \texttt{build}, \texttt{out},
  \texttt{.gradle}, \texttt{.mvn} \\
  Go (additional) &
  \texttt{bin}, \texttt{dist}, \texttt{build} \\
  Rust (additional) &
  \texttt{target}, \texttt{dist}, \texttt{build} \\
  \bottomrule
  \end{tabular}
  \end{table*}

\section{Detailed RQ1 Results}
\label{app:detailed_results}

\begin{figure*}
    \centering
    \includegraphics[width=1\linewidth]{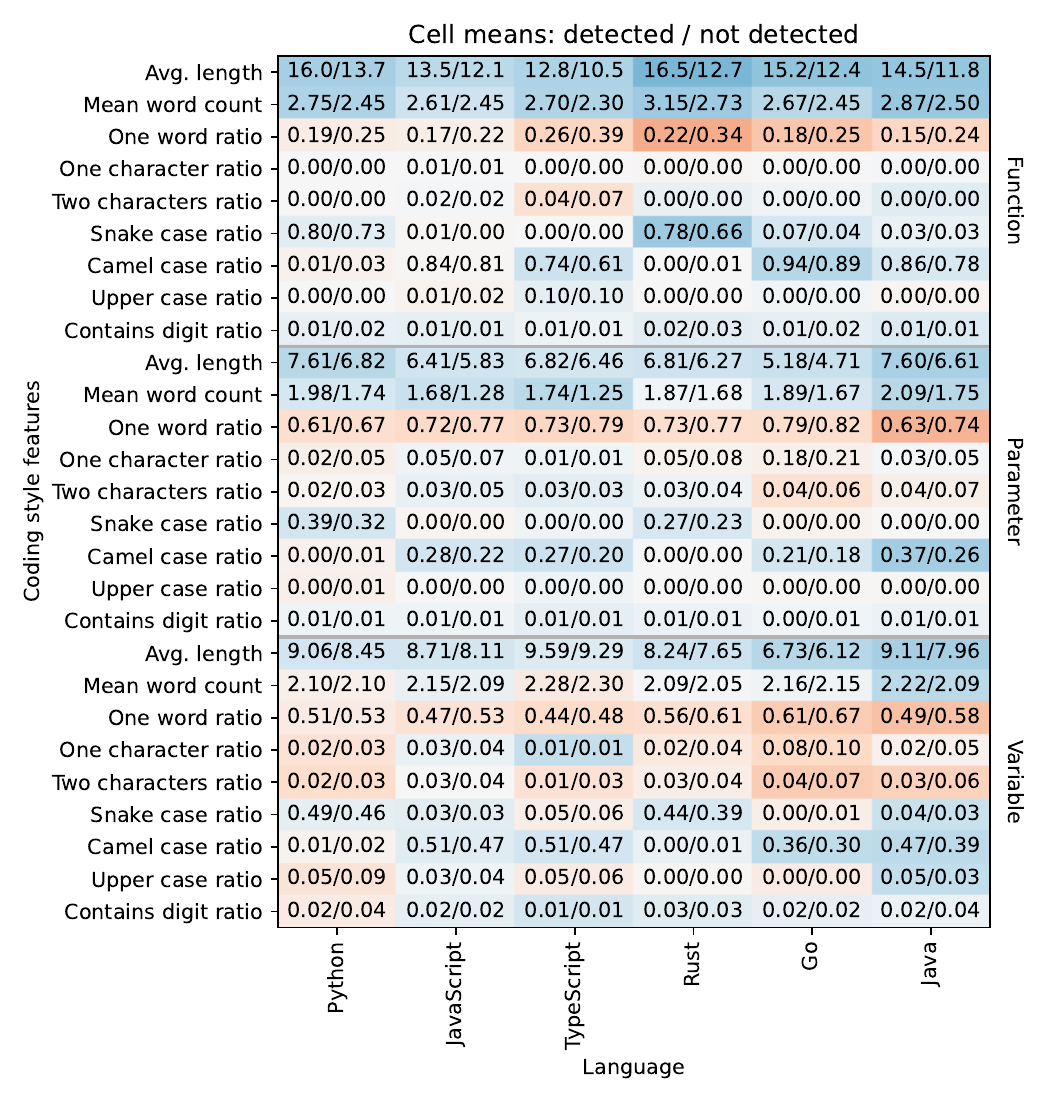}
     \caption{Repository-level naming-feature means for the detected and not-detected groups in RQ1. Each cell reports the detected-group mean followed by the not-detected-group mean, separated by a slash. Background colors encode Mann--Whitney AUC using the same scale as Figure~\ref{fig:llm_related_naming_features}: blue indicates higher values in the detected group, red indicates lower values, and white corresponds to an AUC of 0.5.}
    \label{fig:rq1_group_means}
\end{figure*}

Figure~\ref{fig:rq1_group_means} supplements Figure~\ref{fig:llm_related_naming_features} with group means for all 27 naming features in each language. Means are calculated on the original feature scales, with equal weight per repository.

\section{Detailed RQ3 Results}
\label{app:detailed_rq3_results}
\begin{figure*}
    \centering
    \includegraphics[width=1\linewidth]{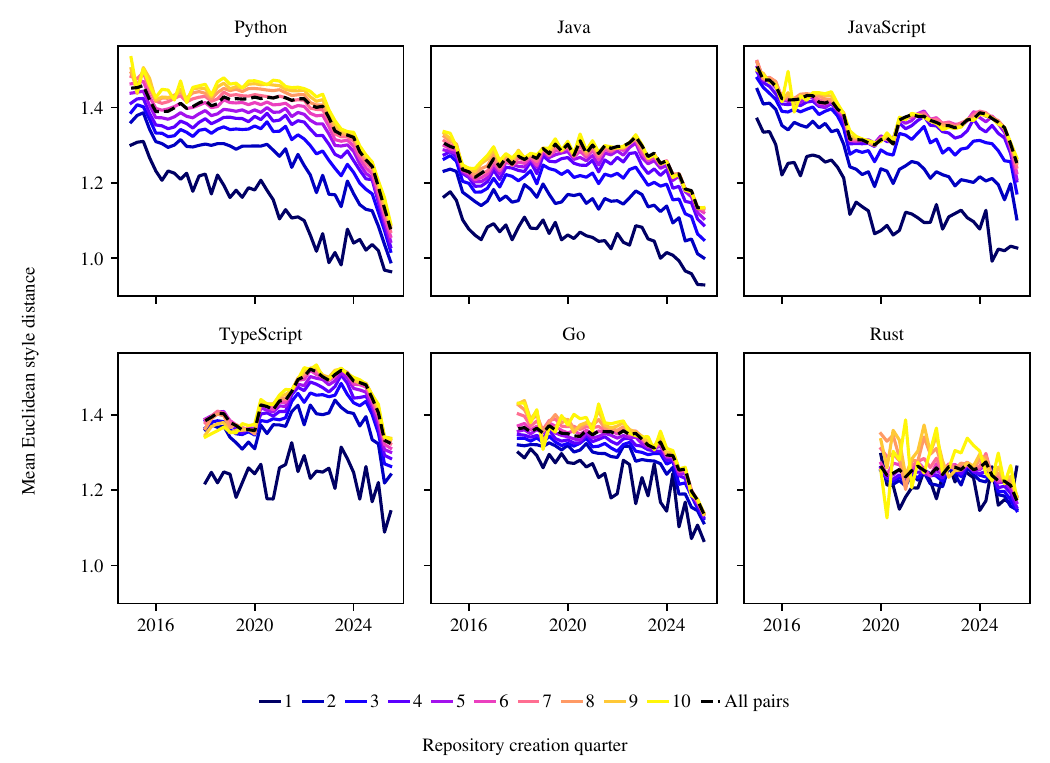}
      \caption{
          Naming-style distance by repository creation quarter
          for owner separations of 1--10 hops in the fixed
          collaboration graph.
          Panels show six programming languages.
          Colored curves show hop-specific distances using the
          owner-balanced aggregation defined in
          Section~\ref{sec:network_locality}.
          Black dashed curves show the equally weighted mean over
          all distinct owner pairs, including separations above
          10 hops.
          Hop-specific points with fewer than 100 contributing
          focal owners are omitted, without excluding those
          owners or pairs from the overall reference.
      }
      \label{fig:network_and_naming_distance_all_hops}
\end{figure*}

Figure~\ref{fig:network_and_naming_distance_all_hops} extends Figure~\ref{fig:network_and_naming_distance} by displaying all hop-specific curves from 1 to 10. The underlying data, aggregation procedure, axis limits, and overall reference are identical to those in the main figure. Hop-specific points with fewer than 100 contributing focal owners are omitted; their coverage is reported in Figure~\ref{fig:rq3_focal_owner_coverage}.

\end{document}